# Oscillation of $T_c$ depending on the number of $CuO_2$ planes in the cuprates

A. Messad

*Laboratoire de Physique, Lycée Olympe de Gouges, 93130 Noisy-Le-Sec, France.*

Using the Tc formula given by BCS, together with a shear modulus varying with the numbering of planes, we can demonstrate that Tc oscillates depending on the number of the cuprate planes. The maximum in these oscillations appears for three planes irrespective of the multilayered cuprate family. Comparison with experimental data shows good agreement and suggests that the electron-phonon coupling predominate in the superconductivity of these cuprates. This model predicts that such oscillations can also occur in many others physical quantities that depend on Tc. The new Tc formula shows interesting prospects towards room temperature superconductors.

### 1. Introduction

The critical temperature of the multilayered cuprates superconductors depends on the number of adjacent cuprate planes in the unit cell. If within a given cuprate family that is taken at the optimal doping, we vary the number n of the cuprate planes, we can observe that Tc follows a curve that looks like a dome whose maximum is at n = 3. This was experimentally well shown for the mercurocuprates family [1]. However, all experimentalists have pointed out the difficulty of elaborating and characterising compounds for which n is greater than 4 or 5 [2].
Recently, some experimentalists who used better materials have come to recognize that Tc is constant or hardly varies when n is superior to 5 or 6 [3].
Using the classic BCS expression of Tc together with a model of the shear modulus of a stacking of cuprate planes, we show that Tc oscillates depending on n and that the famous n = 3 arises naturally as being the localisation of the first maximum in these oscillations.
In this paper, I expound an in-depth analysis of previous results [4] and I show how they can be used to track very high-Tc superconductors.

### 2. Model and Application

According to BCS, Tc is proportional to the Debye temperature. This latter was given by Anderson [5]. So for Tc, we have

$$Tc \propto \left(\frac{h}{k_B}\right) \cdot \left(\frac{3q N_A}{4\pi \Omega_{mol}}\right)^{1/3} \cdot \left(\frac{G}{\rho}\right)^{1/2} \qquad (1)$$

and after some transformations

$$Tc \propto \frac{q^{1/3} \cdot G^{1/2}}{M^{1/2}} = f(n) \cdot G^{1/2} \qquad (2)$$

where *h, $k_B$* and $N_A$ are respectively Planck, Boltzmann and Avogadro constants.

$\Omega_{mol}$ and $\rho$ are respectively the molecular volume and the density of the material.
$G$ is the shear modulus of the crystal.
$M$ is the molecular mass.
$q$ is the number of atoms per molecule.

For a given cuprate family, the function f(n) is obtained from the chemical formula as shown by Table 1.
Supposing that the cuprate planes behave like parallel springs, the shear modulus of the p$^{th}$ plane is given, as expected in metals [6], by

$$G_p \propto \frac{\sin(k_0 a_0 p)}{k_0 a_0 p} \qquad (3)$$

Given that $dc/a_0 = dp$ (where c is the c-axis parameter and $a_0$ is the interplane spacing), the summation over n planes gives the global shear modulus

$$G \propto \int_0^n \frac{\sin(k_0 a_0 p)}{p} dp \propto si(k_0 a_0 n) \qquad (4)$$

where *si* is the *sine integral function* [7] and $k_0$ is a vector of the reciprocal space. We suppose that $k_0 = 1/a_0$ .
Thus the global shear modulus is simply given by

$$G \propto si(n) \qquad (5)$$

Finally, if we suppose that the electron-phonon coupling parameter is a constant within a family [8], we obtain for Tc the following expression

$$Tc \propto f(n) \cdot [si(n)]^{1/2} \qquad (6)$$

In most cuprate families, f(n) hardly varies and thus Tc is reduced to

$$Tc \propto [si(n)]^{1/2} \qquad (7)$$

The plot of Tc versus n (fig.1) shows a set of extrema that are localised at $\pi$, $2\pi$, $3\pi$, etc. The first one is a maximum and appears at exactly $n = \pi$ .

When applying equation (6) under the following form

$$Tc = A \cdot f(n) \cdot [si(K \cdot n)]^{1/2} \qquad (8)$$

to the Hg-family and Tl$_2$-family, we obtain the results of fig.2 and fig.3. The fitting parameters are A and K. The latter is found very close to unity as previously supposed.

## 3. Discussion

When comparing values of A, for points that are far from the curve, with the fitting values, we observe that these points must be 10 to 20 K higher than experimental values. This is probably because the corresponding samples (n = 5 for Hg-family and n = 4 for $Tl_2$-family) were not of good quality. With this correction the fit becomes better.

Values of K are very slightly greater than 1 (1.04 after the above correction) and therefore values of n corresponding to the extrema are 3.02, 6.04, 9.06, 12.08, etc. K is not exactly equal to 1 probably because of a tilt in the cuprate planes.

The first maximum appears for n = 3 as a property of the *sine integral function* and then it is independent from the nature of the material.

If we suppose that the strong electron-phonon coupling occurs within the cuprate planes and that these latter are coupled via electrostatic interactions, then we can show that Tc is approximately given by

$$T_c = B \frac{q^{1/3}}{\varepsilon^{1/2} \times \Omega_{mol}^{1/3} \times M^{1/2} \times \left(e^{2/\lambda} - 1\right)} \qquad (9)$$

where *B* is a constant, $\lambda$ is the electron-phonon parameter and $\varepsilon$ is the dielectric constant. The constant *B* depends, among others things, on the unit cell parameters (probably only on a and b) and on the Poisson's coefficient.

According to (9), a good candidate for very high Tc must contain a great number of light elements of small size and must have small $\varepsilon$ together with high $\lambda$ (i.e., high density of state and small ionic charges). Before a candidate arises from such a compromise, we can already make some predictions with the f(n) formula. Examples of such predictions are presented in Table 2. Some of them seem realistic and others are fictional, but we have to take them into account.

## 4. Conclusion

The critical temperature of the multilayered cuprates oscillates depending on the number of the cuprate planes. This fact is a consequence of a double coupling: firstly of an electron-phonon one in each cuprate plane and secondly of a mechanical one between the cuprate planes. The latter may be occurs via electrostatic interaction in the c-direction.

If this model is correct, then various physical quantities that depend on Tc must oscillate, as for example, the superconducting gap, particularly near Tc.

This model shows that using greater number of lighter elements can achieve higher Tc.

To test completely the validity of this model, more precise experimental measurements on very clean single crystals of higher n will be necessary.

**Acknowledgements:**

I am very grateful to Mrs F. Pillier for providing me with the bibliography. I wish to thank my colleagues C. Martin, H. Ling, M. Herbaut and N. Malguid for reading the manuscript and correcting numerous English mistakes.

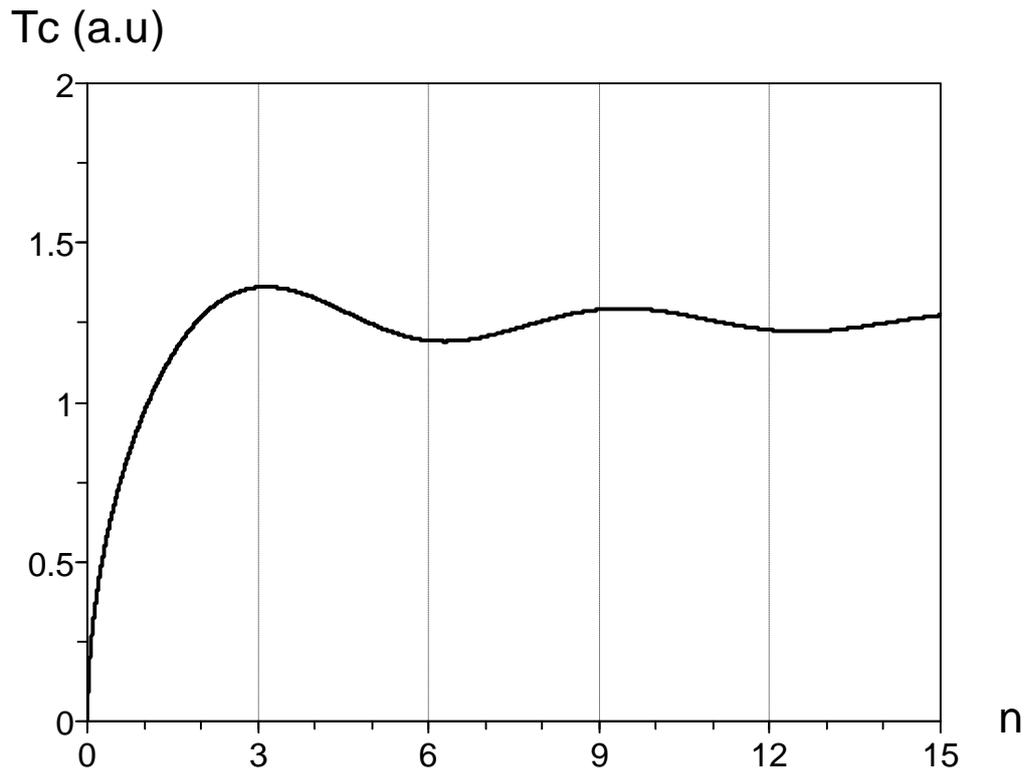

Figure 1: Plot of Tc *vs.* n from Eq. (7).

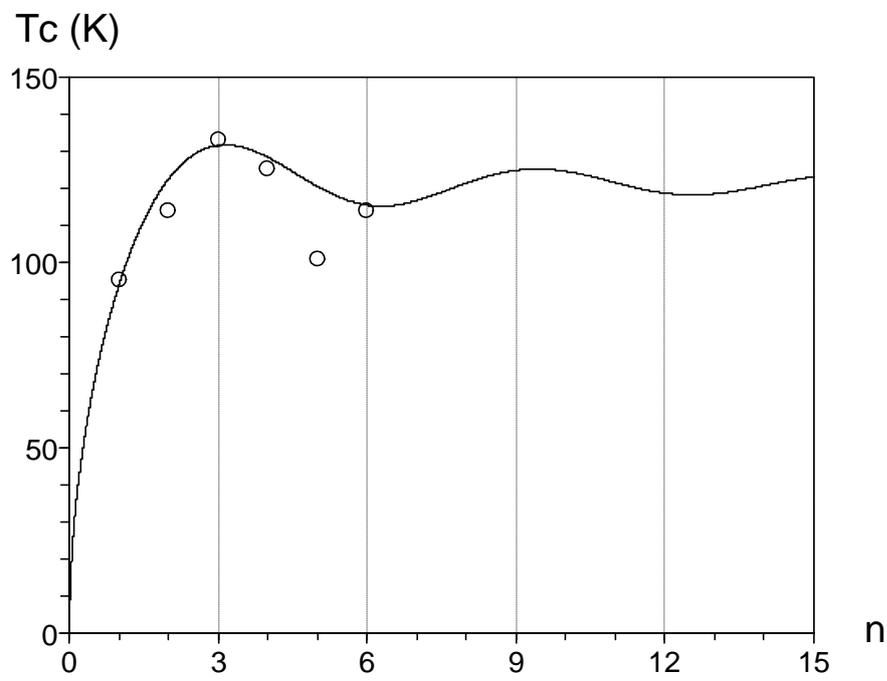

Figure 2: Plot of Tc *vs.* n from Eq. (8). $HgBa_2Ca_{n-1}Cu_nO_{2n+2}$ experimental points (○) were taken from ref. [4].

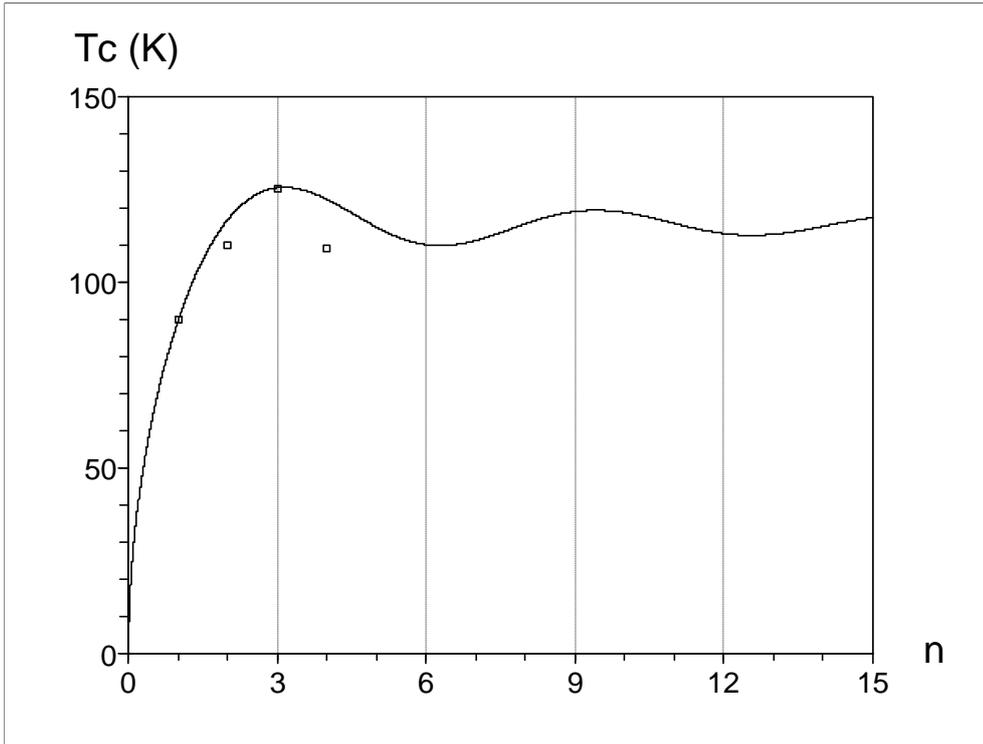

Figure 3: Plot of Tc *vs.* n from Eq. (8). $Tl_2Ba_2Ca_{n-1}Cu_nO_{2n+4}$ experimental points (▫) were taken from ref. [4].

Table 1

| Family | f (n) |
|---|---|
| $TlBa_2Ca_{n-1}Cu_nO_{2n+3}$ | $\dfrac{(4n+5)^{1/3}}{(135.6n+486.9)^{1/2}}$ |
| $Bi_2Sr_2Ca_{n-1}Cu_nO_{2n+4}$ | $\dfrac{(4n+7)^{1/3}}{(135.6n+617.1)^{1/2}}$ |
| $Tl_2Ba_2Ca_{n-1}Cu_nO_{2n+4}$ | $\dfrac{(4n+7)^{1/3}}{(135.6n+707.3)^{1/2}}$ |
| $HgBa_2Ca_{n-1}Cu_nO_{2n+2}$ | $\dfrac{(4n+4)^{1/3}}{(135.6n+467.1)^{1/2}}$ |

Table 2

| Material | Existence | Measured Tc (K) | Predicted Tc (K) |
|---|---|---|---|
| $HgBa_2Ca_2Cu_3O_8$ | real | 135 | - |
| $SrTi_2Ca_2Cu_3O_8$ | imaginary | - | 164 |
| $ZnMg_2Be_2Cu_3O_8$ | imaginary | - | 188 |
| $H_3Ca_2Cu_3O_8$ | imaginary | - | 199 |
| $H_{10}Ca_2Cu_3O_8$ | imaginary | - | 222 |
| $H_{30}Ca_2Cu_3O_8$ | imaginary | - | 268 |
| $H_{50}Ca_2Cu_3O_8$ | imaginary | - | 297 |
| $H_5Cu_3O_8$ | imaginary | - | 222 |
| $H_{15}Cu_3O_8$ | imaginary | - | 257 |
| $H_{35}Cu_3O_8$ | imaginary | - | 302 |
| $H_{145}Cu_3O_8$ | imaginary | - | 396 |